\documentclass[11pt]{article}
\usepackage{amsfonts}

\usepackage{amssymb}
\usepackage{latexsym}
\usepackage{graphicx}
\usepackage[english]{babel}

\begin{document}
\input epsf

\makeatletter
\@addtoreset{equation}{section}
\makeatother


 \begin{center}
{\LARGE Losing information outside the horizon}
\\
\vspace{18mm}
{\bf    Samir D. Mathur
}
\vspace{8mm}

\vspace{8mm}

Department of Physics,\\ The Ohio State University,\\ Columbus,
OH 43210, USA\\ 
\vskip 2 mm
mathur@mps.ohio-state.edu\\
\vspace{10mm}

\end{center}

\thispagestyle{empty}

\def\p{\partial}
\def\h{{1\over 2}}
\def\be{\begin{equation}}
\def\bea{\begin{eqnarray}}
\def\ee{\end{equation}}
\def\eea{\end{eqnarray}}
\def\d{\partial}
\def\la{\lambda}
\def\eps{\epsilon}
\def\bb{\bigskip}
\def\mm{\medskip}
\newcommand{\dm}{\begin{displaymath}}
\newcommand{\edm}{\end{displaymath}}
\renewcommand{\b}{\tilde{B}}
\newcommand{\gm}{\Gamma}
\newcommand{\ac}[2]{\ensuremath{\{ #1, #2 \}}}
\renewcommand{\ell}{l}
\newcommand{\z}{\ell}
\newcommand{\newsection}[1]{\section{#1} \setcounter{equation}{0}}
\def\bb{$\bullet$}
\def\Qbar{{\bar Q}_1}
\def\QPbar{{\bar Q}_p}
\def\q{\quad}
\def\bn{B_\circ}
\def\sq{{1\over \sqrt{2}}}
\def\z{|0\rangle}
\def\o{|1\rangle}
\def\sqi{{1\over \sqrt{2}}}

\let\a=\alpha \let\b=\beta \let\g=\gamma \let\d=\delta \let\e=\epsilon
\let\c=\chi \let\th=\theta  \let\k=\kappa
\let\l=\lambda \let\m=\mu \let\n=\nu \let\x=\xi \let\r=\rho
\let\s=\sigma \let\t=\tau
\let\vp=\varphi \let\vep=\varepsilon
\let\w=\omega      \let\G=\Gamma \let\D=\Delta \let\Th=\Theta
                     \let\P=\Pi \let\S=\Sigma

\def\h{{1\over 2}}
\def\t{\tilde}
\def\r{\rightarrow}
\def\nn{\nonumber\\}
\let\bm=\bibitem
\def\Kt{{\tilde K}}
\def\b{\bigskip}

\let\p=\partial
\def\u{\uparrow}
\def\d{\downarrow}

\begin{abstract}

\b

 Suppose we allow a system to fall freely  from infinity to a point near (but not beyond) the horizon of a black hole. We note that in a sense the information in the system is already lost to an observer at infinity. Once the system is too close  to the horizon it does not have enough energy to send its information back because the information carrying quanta would get redshifted to a point where they get confused with  Hawking radiation. If one attempts to turn the infalling system around and bring it back to infinity for observation then it will experience Unruh radiation from the required acceleration. This radiation can excite the bits in the system carrying the information, thus reducing the fidelity of this information. We find the radius where the information is essentially lost in this way, noting that this radius depends on the energy gap (and coupling) of the system.  We look for some universality by using the highly degenerate BPS ground states of a quantum gravity theory (string theory) as our information storage device.  For such systems one finds that the critical distance to the horizon set by Unruh radiation is the geometric mean of the black hole radius and the radius of the extremal hole with quantum numbers of the BPS bound state. Overall, the results suggest that information in  gravity theories should be regarded not as a quantity contained in a system, but in terms of how much of  this information is accessible to another observer. 

\end{abstract}
\vskip 1.0 true in

\newpage
\renewcommand{\theequation}{\arabic{section}.\arabic{equation}}

\def\p{\partial}
\def\r{\rightarrow}
\def\h{{1\over 2}}
\def\b{\bigskip}

\def\nn{\nonumber\\ }

\section{Introduction}
\label{intr}\setcounter{equation}{0}

If an object falls through the horizon of a black hole then one seems to lose its information, at least from the viewpoint of an observer at infinity. 
But in some considerations of black hole physics, one argues that the physics outside the horizon is complete by itself, and one need never think about the interior of the hole. The idea of black hole complementarity is one such situation, where one assumes that the infalling observer would get destroyed at the horizon (and have his information re-radiated to infinity); it is only in a second complementary description that he falls through the horizon \cite{thooft1,suss1}. Some have argued that matter never falls into black holes because of its ever increasing redshift \cite{sonego}, or that the backreaction of Hawking radiation may be severe enough to prevent a shell from falling through its horizon \cite{vachas}. In string theory one finds that black hole microstates do not have regular horizons; instead they are `fuzzballs' \cite{fuzz}. 

In such situations, one may think that the information in an infalling bit would be preserved until the infalling object reaches the horizon (or the surface of the fuzzball). But as we will note below, the information in the bit is essentially lost to the observer at infinity {\it before} the system  reaches the horizon. We perform the following computations to support this observation:

\b

(a) Suppose the infalling observer tries to send his information out to infinity before he crosses the horizon. If he has fallen to a point $\bar r$ that is close to the horizon, then he is travelling very fast inwards, and will thus need to emit a very energetic photon backwards in order to carry the needed information to infinity. This photon will be redshifted as it climbs out of the potential well at $\bar r$, and reaches infinity with a low energy $E_\gamma$. If $E_\gamma\lesssim T$, where $T$ is the temperature of the hole, then we cannot decode the data in the photon because it cannot be distinguished from the bath of Hawking radiation \cite{hawking} being emitted by the hole. This requirement gives us a minimum mass $m$ that the infalling observer must possess in order to send reliable information out to infinity. 
Equivalently, for a given mass $m$ available to the observer, there is a critical value of the radius $\bar r$ beyond which he cannot reliably send his information out to infinity, even though he has not crossed the horizon.

\b

(b) The above was a very simple observation,  and an immediate objection to this line of thought would be the following. Instead of having the infalling observer send his information out, we could just have the observer turn back (or be pulled back with a rope), and then we could observe the data in his bit back at infinity. But if the infalling observer has reached a position $\bar r$ then there is a minimum acceleration which he needs to maintain (for a certain time) if he is to avoid falling into the hole. This acceleration creates Unruh radiation \cite{unruh}, which will interact with his information carrying bit and (with some probability) change its pure state to a state entangled with the radiation field. Thus we again fail to recover the information in the bit for the purposes of the observer at infinity. We find the critical distance $d_{cr}$ from the horizon where the information is effectively lost this way; the result is
\be
d_{cr}\sim \sqrt{r_H\over \Delta E}
\label{qeight}
\ee
where $r_H$ is the radius of the hole and $\Delta E$ is the energy gap in the 2-level system that we use to model our infalling observer. If we set the mass $m$ in the estimate discussed in (a) to be $m\sim \Delta E$, then we find that the critical distances in the two cases are of the same order.

\b

(c) In the above computation the probability that we flip the bit by Unruh radiation depends on the  energy gap $\Delta E$ between the states of the 2-level system. (In principle we also have a dependence on the coupling $\alpha$ between the system and the radiation field, but in writing (\ref{qeight}) we have fixed $\alpha$ and then taken the limit of close approach to the horizon; in this case the result becomes insensitive to $\alpha$.)  Such a dependence on the parameters of the system is of course a feature of all computations with accelerated detectors, but we would like to see if there are situations where one might obtain some universal results. To do this we should make our system from objects present in a full theory of quantum gravity, which we take to be string theory. One way to store a large amount of information with a low cost in energy is to use  BPS brane bound states, which have  a large degeneracy $N$ for a given value of charges.  If we put the system in one of these states then (in the absence of any disturbance) it would continue to be in that state, storing $\ln N$ bits of information.  

To find the effect of acceleration on such a system, we first find the  probability of excitation for a system with a dense set of excitation levels, relating the result to the absorption cross section $\sigma$ for the system. Next, we note that for extremal brane bound states in string theory we get $\sigma=A_{ex}$,  where $A_{ex}$ is the  horizon area  of the extremal black hole that would have the mass and charges carried by the brane bound state.  We then use these results in the computation of (b) above,  to find the closest distance that our system can approach a large neutral hole and still be pulled back without creating an excitation on the system. We find that this critical distance from the horizon is 
\be
d_{cr}\sim \sqrt{r_H \, r_{ex}}
\ee
where $r_H$ is again the radius of the black hole and $r_{ex}$ is the radius of the extremal hole that carries the quantum numbers of the brane system. 

\b

In short, one finds that there are limitations on how well information can be preserved once a system falls into a strong gravitational potential. The computations of this paper indicate a few of the possible constraints; more exploration would be needed to see if there are universal expressions relating the energy available for encoding information, the information stored, and the depth of the gravitational potential well.

Many related directions have been explored in relating gravity and information.  In \cite{degrade, degrade2} an accelerated bit was studied, and its state was seen to entangle with the field by an amount depending on the acceleration $a$. This computation is thus in the same spirit as the ones we are interested in, but our interest is in how close to a black hole a system  can fall before the its information becomes hard to recover.   Accelerated detectors and detectors falling into black holes have limits on what they can observe \cite{hayden}. If we accelerate one member of an entangled pair then the entanglement is altered \cite{louko,benatti,mann}. Hawking radiation effects affect entanglement as well \cite{ahn}. The recoil of an accelerated detector emitting emitting radiation decoheres the emitted fluxes \cite{parentani}. There has been considerable work on the validity of the second law in the context of black holes \cite{unruhwald}.  General reviews of entropy, black holes and information flow can be found in \cite{bekreview,peres}.
A good review of the Unruh effect is given in \cite{crispino}. Acceleration radiation has also been studied in the holographic context \cite{das}.

\section{Emitting information back to infinity}

\begin{figure}[htbp]
\begin{center}
\includegraphics[scale=.38]{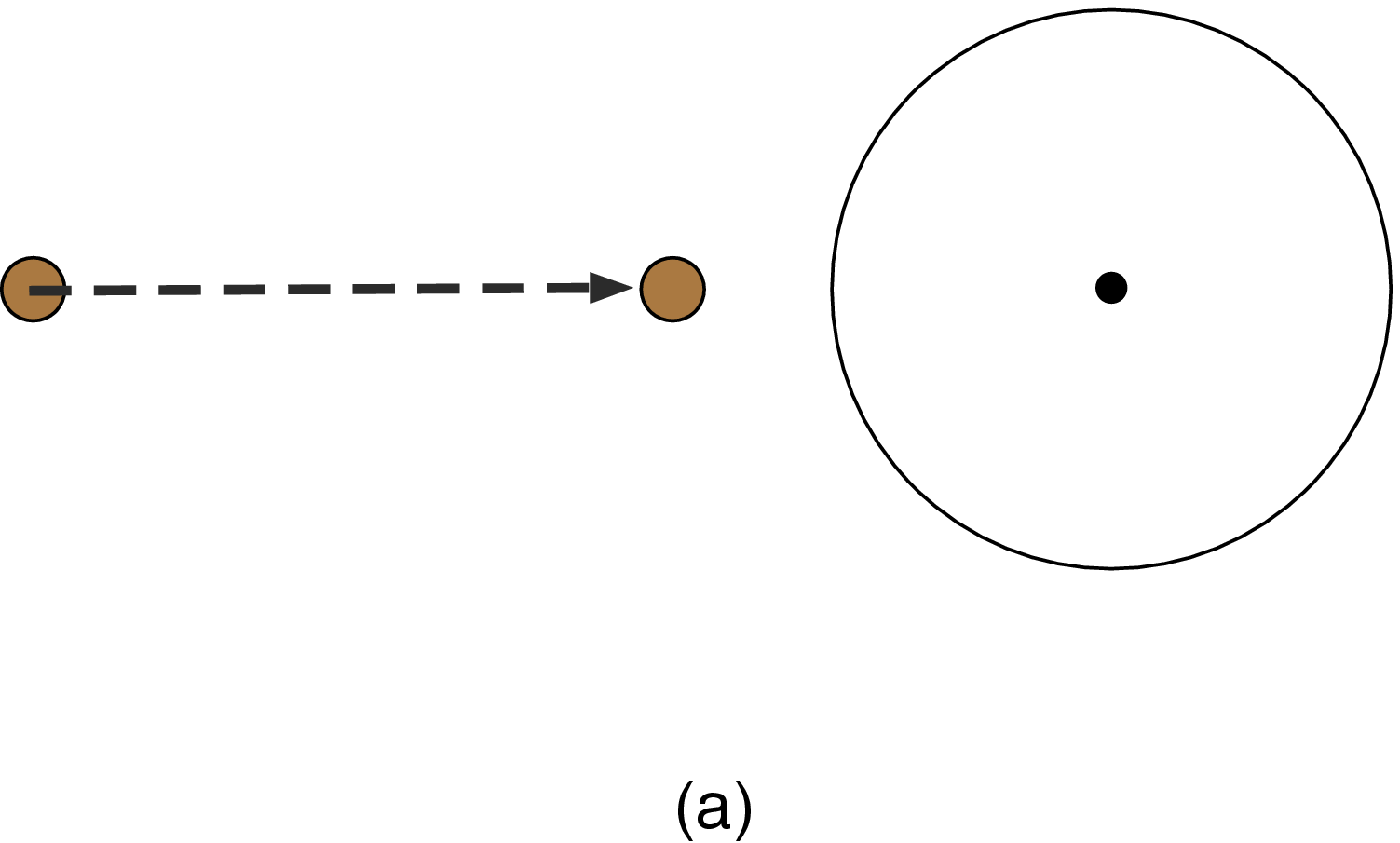}\hskip 1 true in
\includegraphics[scale=.38]{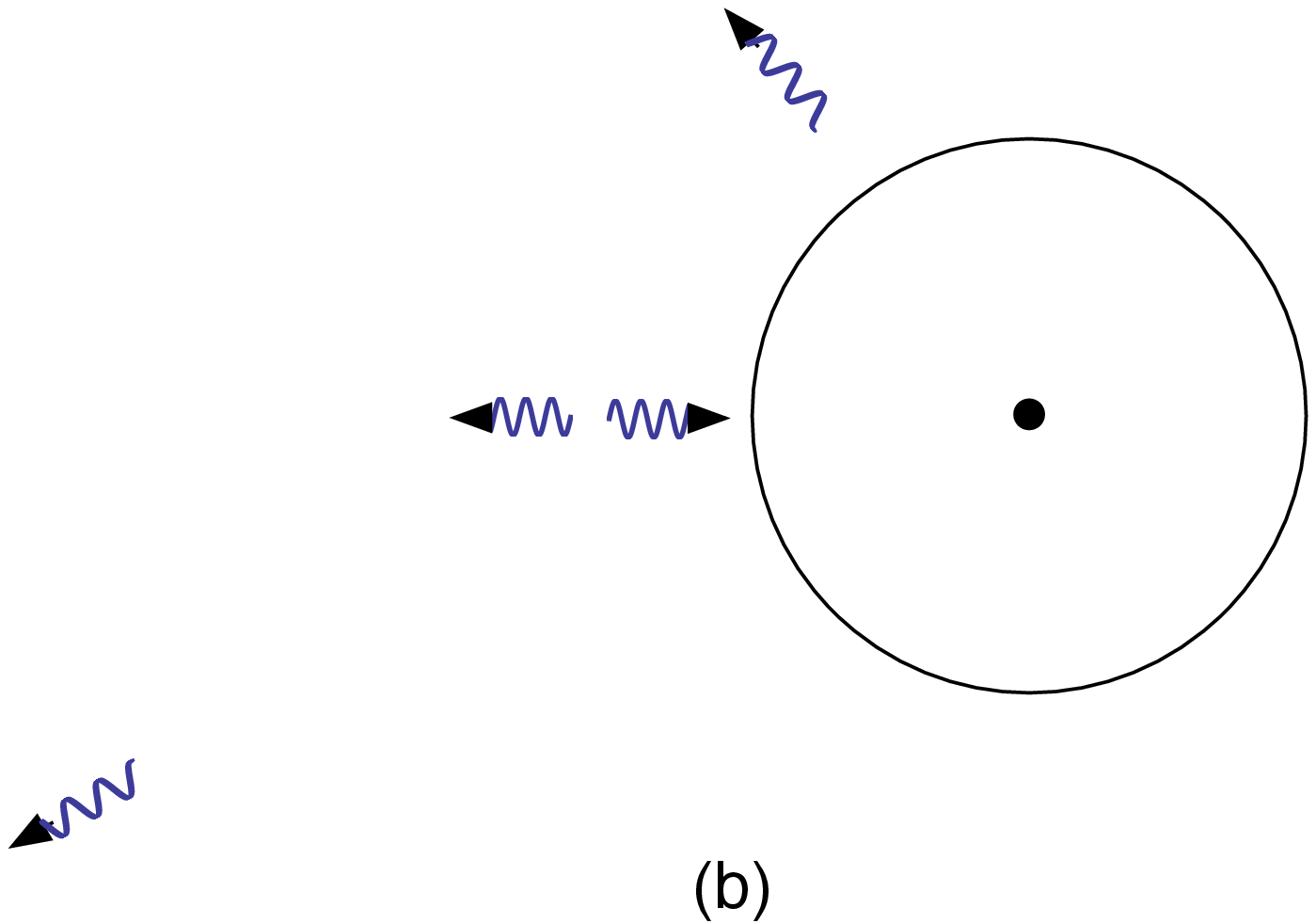}
\caption{{(a) A system of mass $m$ falls from rest at infinity to a point $\bar r$ close to the horizon $r_H$ (b) The system breaks into two parts so that it can send information back using a high energy photon, but this photon must have more energy at infinity than Hawking quanta in order to be distinguished from them. }}
\label{foneftwo}
\end{center}
\end{figure}

Let us begin by examining the situation (a) listed in the introduction. Fig.\ref{foneftwo}(a) shows a system falling from rest at infinity to a point $\bar r$ near the horizon. We assume that this system carries some information, which it would like to send back out to infinity. Thus at the point $\bar r$ the system emits an outgoing photon, as shown in fig.\ref{foneftwo}(b). By momentum conservation, the rest of the system will get an inward kick. The total energy of the system is $m$, its rest mass. A simple computation shows that the energy of the outgoing photon is maximized when the ingoing part is also massless; in this case each of the two halves gets an energy ${m\over 2}$ in the local Lorentz frame moving with the infalling system.

The outgoing photon will get redshifted by the time it reaches infinity, reaching infinity with some energy $E_\gamma$. If the information in this photon is to be useful, then we must be able to differentiate this photon from the photons emitted by the hole as Hawking radiation. Thus we need $E_\gamma\gtrsim T$, where $T$ is the temperature of the hole. 

This requirement sets a limit on how close to the hole the system can fall before it attempts to send its information back out. Let us compute this location in more detail.

We will consider black holes in $D$ spacetime dimensions with metric
\be
ds^2=-f dt^2+ {dr^2\over f}+r^2 d\Omega_{D-2}^2
\label{three}
\ee
where
\be
f=1-({r_H\over r})^{D-3}
\ee
Consider a particle of mass $m$ that falls in radially, starting from rest at infinity. Let $U^\mu={d x^\mu\over d\tau}$ be the proper velocity of the infalling particle; the nonzero components are $U^t, U^r$. These components can be determined by the normalization condition
\be
-f(U^t)^2+{1\over f}(U^r)^2=-1
\label{one}
\ee
and the conservation of energy
\be
E=-p_t=-m g_{tt} U^t
\label{two}
\ee
Since we have taken the particle to fall from rest at infinity, we have $E=m$. From (\ref{one}) and (\ref{two}) we get
\be
U^t={1\over f}, ~~~U^r=-\sqrt{1-f}
\label{four}
\ee

When this particle falls to a radius $\bar r$ we assume that it splits into two parts, with one part being a radially outgoing photon with energy $E_\gamma$. Let $\hat E_\gamma$ denote the energy of this photon in the local Lorentz frame moving with the infalling system at $\bar r$. 
Using energy and momentum conservation, we find that the largest $\hat E_\gamma$ is attained when the infalling part is also massless; in this case we get
\be
\hat E_\gamma=\hat p_\gamma={m\over 2}
\ee
In the coordinates (\ref{three}), the condition $p_{\gamma a}p^a_\gamma=0$ for the massless outgoing quantum gives
\be
p^r_\gamma=f p^t_\gamma
\label{five}
\ee
We have
\be
\hat E_\gamma=-p_{\gamma a}U^a={p^t_\gamma}[1+\sqrt{1-f(\bar r)}]
\ee
where we have used (\ref{four}),(\ref{five}).
Setting $\hat E_\gamma$ to ${m\over 2}$, we find for the conserved energy of the outgoing massless quantum
\be
E_\gamma=-p_{\gamma t}=f(\bar r)p_\gamma^t={mf(\bar r)\over 2[1+\sqrt{1-f(\bar r)}]}\approx {m\over 4} f(\bar r)
\ee
where the approximation in the last step holds for $\bar r$ close to $r_H$. Let us require that $E_\gamma$ equal $T$, where $T$ is the temperature of the hole. Then we get ${m\over 4} f(\bar r)\approx T$, which gives
\be
{\bar r-r_H\over r_H}\approx {4T\over (D-3) m}
\label{foone}
\ee
Noting that $T={(D-3)\over 4\pi r_H}$, we see that
\be
{\bar r-r_H}\sim {1\over m}
\label{thfour}
\ee

As an example, consider the Schwarzschild hole in 3+1 dimensions with mass $M$. If we take $m=20 T$, then (\ref{foone}) gives $r-r_H\approx 0.4 M$.
With a little more effort we could investigate the fidelity of the returned information sent as a function of $m$; we hope to return to such an investigation elsewhere.

\section{Unruh radiation felt by a returning system}

Now we consider the process (b) discussed in the introduction. We saw above that once a system falls too close to the horizon, it does not have enough energy to send out information reliably. But since the observer at infinity has an arbitrary amount of energy at his disposal, it would seem that he can  use an external device to pull the infalling system back to infinity, and then decode the state of the system at leisure.

But this method of retrieving information also meets with a difficulty. The infalling system gathers a large inward velocity by the time it falls close to the horizon. To turn it back and bring it to infinity we have to provide an acceleration, and thus any information carrying bits in the system will interact with the Unruh radiation resulting from this acceleration. The near horizon region of the black hole is just Rindler space, so we can study the kinematics of accelerating observers in the simpler setting of Rindler space. We observe that if we provide too low an acceleration, the system will just continue with its large inward velocity and fall past the horizon. If we provide a very large acceleration, then we will be able to extract the system back to infinity, but the large acceleration will very likely change the state of the information carrying bit  due to the interaction with Unruh radiation. 

We should therefore proceed in the following steps. In (\ref{four}) we have already found the radial velocity attained by the system as it falls to a position $\bar r$, starting from rest at infinity. Suppose at this point we decide to start bringing the system back. We switch to local Rindler coordinates for the near horizon geometry to make the computation simpler. There are an infinite number of possible paths that we could use, and in principle we should investigate all of them and find which one leads to the smallest probability $P$ for excitation of the system. While this could be done in principle (perhaps numerically), here we will content ourselves with looking at a 1-parameter family of paths, which maintain a constant acceleration $a$ from the time the particle is at the radius $\bar r$ (with proper velocity $\bar U^r$), to the time when we bring it back to $\bar r$  with the {\it opposite} velocity $-\bar U^r$. We then let the system escape back to infinity, in a reverse of the initial free fall motion. The free parameter is the acceleration $a$. The value of $a$ determines the proper time $\Delta \tau$ for which we will have to maintain the acceleration. The probability of excitation is $P=\Gamma\Delta \tau$ where the probability of excitation per unit time $\Gamma$ is given in terms of $a$ by the standard computation of the excitation rate for accelerating detectors.

The limit $a\r \infty$ describes a sudden reflection of the radial velocity; in this limit $\Delta \tau$  will go to zero but the net effect of the large acceleration may disturb the system significantly. If $a$ is chosen too small, the system will just fall through the horizon. If we choose $a$ to be just large enough that the system does not fall in, then the system will  stand near the horizon with constant acceleration for a diverging amount of time, and again the system will be significantly affected. Thus we have to find the optimal acceleration $a$ (and the corresponding time $\Delta \tau$), which gives the minimal probability of excitation for the system.

We let the infalling system be a 2-level system, where the  two states are separated by an energy gap $\Delta E$. We can store one bit on information in this system, by choosing it to be in one of the two states.   We assume that this system is coupled (with a coupling constant $g$) to a massless scalar field.  We start with the system in the lower energy state, and compute the  probability that the system gets excited to the upper state; if this happens then we have lost the information we had tried to keep in the system. 

For $r\approx r_H$ the metric (\ref{three}) describes Rindler spacetime
\be
 ds^2=- r_R^2 \, dt_R^2+dr_R^2+dz_R^idz_R^i
 \label{eight}
 \ee
 where we have defined
 \be
 t_R={D-3\over 2 r_H} t, ~~~r_R=\sqrt{4r_H(r-r_H)\over D-3}
 \label{seven}
\ee
and $z_R^i, i=1, \dots D-2$ are the directions along the horizon surface.

\b

{\it Notation:} We have used coordinates $t, r$ etc for the original $D$ dimensional metric (\ref{three}). In the near horizon Rindler frame (\ref{eight}) we will let all variables carry a subscript $R$. The accelerated path segment begins and ends at $\bar r$, and we will add a bar to all variables to denote their values at these endpoints. Thus $\bar U^r_R={d r_R\over d\tau}(\bar r_R)$ is the proper velocity in the Rindler frame at the point $\bar r_R$ (we will pass through the point $\bar r_R$ twice, once when falling in and once returning, so we have to be careful to also specify the sign of $\bar U^r_R$). Finally, the acceleration in Rindler space will be written as
\be
a\equiv{1\over r_{1R}}
\ee
since the variable $r_{1R}$ will be more convenient than $a$.

\subsection{The path of the system in Rindler space}

\begin{figure}[htbp]
\begin{center}
\includegraphics[scale=.38]{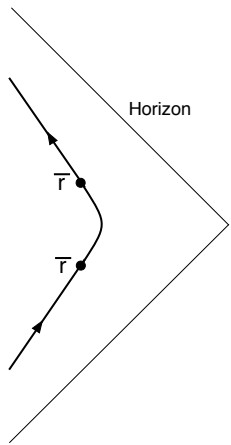}
\caption{{The trajectory of the system in the $t-r$ plane. We have free infall until the position $\bar r$, then a period $\Delta\tau$ of constant acceleration $a$,  then free motion again out to infinity (along the reverse of the infall trajectory).}}
\label{fthree}
\end{center}
\end{figure}

We should first find the kinematics of the paths we consider.
The system will be in free fall until the radius $\bar r$. This will be a point $\bar r_R$ in the Rindler coordinates (\ref{seven}). The radial velocity $\bar U^r$ at $\bar r$ will take a value $\bar U^r_R$ in Rindler coordinates. The path of our system is described as follows  (fig.\ref{fthree}):

\b

(i)  We have free fall to the point $\bar r_R$ (as mentioned above). At this point the radial velocity is $U^r_R(\bar r_R)=-|U^r_R(\bar r_R)|$

\b

(ii) We have a period of constant acceleration, with proper acceleration $a$, for a period of proper time $\Delta \tau$. This brings the system back to the location $\bar r_R$, with a velocity $U^r_R(\bar r_R)=|U^r_R(\bar r_R)|$

\b

(iii) The system flies (with no acceleration) along the reverse of the path in (i), reaching infinity with no velocity. 

\b

We now need to find the path with these properties in the Rindler coordinates (\ref{eight}). 

\subsubsection{Paths of constant acceleration}

We  first write down the paths with constant acceleration in Rindler space. (We will not write down the transverse directions $z_R^i$ in (\ref{eight}) in the following discussion.) Rindler spacetime is one quadrant of Minkowski spacetime; the latter is described by coordinates $T,X$ with
\be
T=r_R\sinh t_R, ~~~X=r_R\cosh t_R
\ee
A path at constant $r_R=r_{R1}$ has a constant acceleration 
\be
a={1\over r_{R1}}
\label{tsix}
\ee
 Such a path will be described by
\be
T=r_{R1}\sinh {\tau\over r_{R1}}, ~~~X=r_{R1}\cosh {\tau\over r_{R1}}
\label{tenp}
\ee
where $\tau$ is the proper time along the path.\footnote{The proper time $\tau$ does not need a subscript $R$ since it is the same in all frames.} But constant $r_R$ paths are not the only trajectories with constant acceleration. The most general constant acceleration trajectory is given by shifting the constant $r_R$ path by a constant change in the Minkowski coordinate $X$
\be
T=r_{R1}\sinh {\tau\over r_{R1}}, ~~~X=r_{R1}\cosh {\tau\over r_{R1}}+b
\label{ten}
\ee
(We will always shift $T$ so that the point of minimum $r_R$ occurs at $T=0$.) The acceleration is still $a={1\over r_{R1}}$. The  minimum value of $r_R$ along the path is
\be
r_{R, min}=r_{R1}+b
\label{fourt}
\ee

\subsubsection{Finding the constant $b$}

We wish to find a path of the form (\ref{ten}) that will pass through $\bar r_R$ with radial velocities $\pm |\bar U^r_R|$. The acceleration $a=1/r_{R1}$ is also chosen by us. Given these constraints, we wish to find $b$ and thus determine the path followed by our system.

We have
 \be
r_R^2=X^2-T^2=r_{R1}^2+b^2+2b\,r_{R1} \cosh{\tau\over r_{R1}}
\ee
Differentiating this gives $2 r_R {dr_R\over d\tau}=2b\sinh{\tau\over r_{R1}}$. Thus we find
\be
U^r_R(\bar r_R)=\pm {b\over \bar r_R}\sinh{\bar \tau\over r_{R1}}\equiv \pm\bar U^r_R
\label{thir}
\ee
where $\pm \bar\tau$ is the value of  the proper time at the initial and final points of the accelerating path segment. 

Eq. (\ref{thir}) gives one relation that relates $b$ to the constraints on the path. Next, we write (\ref{ten}) by expressing $T,R$ in Rindler coordinates
\bea
r_{R1}\sinh {\tau\over r_{R1}}&=&r_R\sinh t_R\cr
r_{R1}\cosh {\tau\over r_{R1}}+b&=&r_R\cosh t_R
\label{el2}
\eea
The first of these equations gives 
\be
\sinh \bar t_R={r_{R1}\over \bar r_R}\sinh {\bar\tau\over r_{R1}}={r_{R1}\bar U^r_R\over b}
\ee
The second then gives
\be
\bar r_R \cosh \bar t_R=\bar r_R \sqrt{1+({r_{R1} \bar U^r_R\over b})^2}=r_{R1}\cosh{\bar\tau\over r_{R1}}+b=r_{R1}\sqrt{1+({\bar r_R\bar U^r_R \over b})^2}+b
\ee
which is
\be
\bar r_R \sqrt{b^2+({r_{R1}\bar  U^r_R})^2}=r_{R1}\sqrt{b^2+({\bar r_R\bar U^r_R })^2}+b^2
\label{tw}
\ee
 This has the solutions (note that the squareroots in (\ref{tw}) can take either sign)
\be
b=0, ~~b=[r_{R1}^2+\bar r_R^2+2r_{R1}\bar r_R \sqrt{1+(\bar U^r_R)^2}]^\h, ~~b=[r_{R1}^2+\bar r_R^2-2r_{R1}\bar r_R \sqrt{1+(\bar U^r_R)^2}]^\h
\ee
The first solution gives a divergent $\bar\tau$ from (\ref{thir}), so we discard it. The second solution gives $b>r_{R1}+\bar r_R$. But we need to have the minimum value of $r_R$  to satisfy $0<r_{R, min}<\bar r_R$ (since $r_R=0$ is the horizon, and $\bar r_R$ is the point from which we started the accelerating segment of the path.) From (\ref{fourt}) we find
\be
0<r_{R1}+b<\bar r_R ~~~\Rightarrow~~~- r_{R1}<b<\bar r_R-r_{R1}
\ee
Thus we cannot have the second solution, and we take
\be
b=[r_{R1}^2+\bar r_R^2-2r_{R1}\bar r \sqrt{1+(\bar U^r_R)^2}]^\h
\ee
This determines the path, with acceleration $a=1/r_{R1}$, passing through the point $\bar r_R$, with proper velocities $\pm \bar U^r_R$. 

\subsubsection{The proper time of acceleration $\Delta \tau$}

Finally, from (\ref{thir}) we find that the proper time along the accelerating segment is
\be
\Delta \tau=2\bar\tau=2 r_{R1} \sinh^{-1} [{\bar r_R \bar U^r_R\over [r_{R1}^2+\bar r_R^2-2r_{R1}\bar r_R \sqrt{1+(\bar U^r_R)^2}]^\h}]
\label{tseven}
\ee
In this expression we should substitute the value of $\bar U^r_R$, which is determined by the fact that we have free fall from rest at infinity to the point $\bar r_R$. In the coordinates (\ref{three}) we had $|\bar U^r|=\sqrt{1-f}$ (eq. (\ref{four})). Converting to the Rindler frame using (\ref{seven}) we find
\be
\bar U^r_R=\sqrt{r_H\over (D-3)(\bar r-r_H)}\sqrt{1-f}\approx \sqrt{r_H\over (D-3)(\bar r-r_H)}\approx {2r_H\over (D-3)\bar r_R}
\label{teight}
\ee
where we have used the approximations valid for $\bar r$ close to $r_H$.

\subsection{The probability of excitation}

The probability of exciting the system from the lower level to the upper level is given by
\be
P=\Gamma \Delta \tau
\label{fift}
\ee
where $\Gamma$ is the probability for excitation per unit proper time along the accelerated path and $\Delta \tau$ is the time for which the system accelerates. In the usual computations of accelerated detectors one assumes that the acceleration persists for infinite time, and the rate $\Gamma$ is computed under this assumption. We on the other hand have acceleration only for a time $\Delta \tau$, and for the relation (\ref{fift}) to make sense we should have
\be
\Delta \tau \gg {1\over a}
\label{sixt}
\ee
i.e., the acceleration should persist for a time much longer than the time scale set by the acceleration itself. But we can see that (\ref{sixt}) does hold in our physical problem, by performing a rough estimate of scales.  Consider flat Minkowski space, and a particle which reverses its trajectory from $U^\mu=(\gamma, v\gamma, 0, 0, \dots 0)$ to $U^\mu=(\gamma, -v \gamma, 0, 0, \dots 0)$. Since ${d U^\mu\over d\tau}=a^\mu$, we have  $\delta U^\mu\sim a^\mu \Delta\tau$, and  $\delta U^\mu\delta U_\mu \sim a^2 (\Delta\tau)^2$. But $\delta U^\mu\delta U_\mu=4 v^2\gamma^2$, so we get $a\Delta \tau\sim 2v\gamma$. Thus if the boost factor $\gamma$ is much larger than unity (i.e. we reflect a fast moving particle) then (\ref{sixt}) will be satisfied. 

In our problem we can compute $\delta U^\mu \delta U_\mu$ as follows. We write $f(\bar r)\equiv \bar f$. From (\ref{four}) we see that at the start of the acceleration we have $U^\mu=({1\over \bar f}, -\sqrt{1-\bar f}, 0, 0\dots 0)\equiv U^\mu_i$. At the end of the acceleration we have $U^\mu=({1\over \bar f}, \sqrt{1-\bar f}, 0, 0, \dots 0)\equiv U^\mu_f$.  
Then
\be
\delta U^\mu \delta U_\mu=(U^\mu_f-U^\mu_i)(U_{\mu, f}-U_{\mu, i})=-2-2U^\mu_f U_{\mu, i}=-4+{4\over \bar f}
\ee
When we fall near the horizon ($r\approx r_H$) then $\bar f\r 0$, and we see that $\delta U^\mu \delta U_\mu\gg 1$. So we see that particles that fall close to the horizon and then accelerate to escape will have an acceleration that satisfies (\ref{sixt}). We can thus use the traditional computation of the excitation rate $\Gamma$ and multiply by $\Delta \tau$ to get $P$. 

\subsubsection{Rate of excitation for an accelerated detector}

Let us now compute $\Gamma$, the probability of excitation per unit time for an accelerated detector.  This is a standard computation, but we outline the steps here so that we get the result in the notation that we have used. 

The paths (\ref{ten}) are shifted versions of the basic path (\ref{tenp}), and thus have the same acceleration. Thus to find the rate of excitation we can just focus on the constant $r_R$ paths (\ref{tenp}) and find the excitation rate as a function of the location $r_R=r_{R1}$ of the detector.

Let the lower energy state of the 2-level system be $|\psi_i\rangle$ and the upper energy state be $|\psi_f\rangle$. The system interacts with a massless scalar field $\phi$ through the interaction Hamiltonian $H_{int}=g\hat O \hat \phi$, where the operator $\hat O$ acts on the 2-level system. We write
\be
g\langle\psi_f | \hat O |\psi_i\rangle\equiv \alpha
\label{thfive}
\ee
The scalar field can be expanded in Rindler modes as
\be
\hat \phi = \sum_{\omega, k_\perp} [{1\over \sqrt{2\omega}}e^{-i \omega t_R} e^{ik_\perp\cdot z_R}F_{\omega, k_\perp}(r_R) \hat a_{\omega, k_\perp}
+ {1\over \sqrt{2\omega}}e^{i\omega t_R} e^{-i k_\perp\cdot z_R}F^*_{\omega, k_\perp}(r_R) \hat a^\dagger_{\omega,k_\perp}]
\ee
where $z^R_i$ are the transverse coordinates in (\ref{eight}). The functions $F$ satisfy the equation
\be
{\p^2 F\over \p r_R^2}+{1\over r_R} {\p F\over \p r_R}+[{\omega^2\over r_R^2}-k_\perp^2]F=0
\ee
and are thus given by \cite{fulling}
\be
F_{\omega, k_\perp}(r_R)={\sqrt{2\omega}\over \pi}\sqrt{\sinh[\pi \omega]} K_{i\omega}(k_\perp r_R)
\ee
where $K$ is the modified Bessel function. 
The thermal state in each Rindler mode is given by a density matrix of the form
\be
\rho=\sum_{n=0}^\infty e^{-\h \beta n\omega}|n\rangle\langle n|=\sum_{n=0}^\infty e^{-\pi n\omega}|n\rangle\langle n|
\label{twenty}
\ee
where we have used the Rindler temperature given by $\beta=T^{-1}={ 2\pi}$ from the  metric (\ref{eight}).

Let the system be in the lower energy state $|\psi_i\rangle$ before the acceleration. Consider the interaction with the  mode $(\omega, k_\perp)$ which leads to the system getting excited to the state $|\psi_f\rangle$. 
The 2-level system moves along the path 
\be
r_R={\rm constant}=r_{R1}
\ee
The proper time along the path is $\tau=r_{R1}{t_R}$. The state $|\psi_i\rangle$ evolves as $e^{-i E_i  \tau}$ from proper time  $-\bar \tau$ until  $\tau$, at which point the system is excited to the state $|\psi_f\rangle$ which evolves as $e^{-i E_2  \tau}$ until the time $\bar\tau$. If the scalar mode is in the state $|n\rangle$ before the acceleration, then after the acceleration we get
\be
|n\rangle\otimes |\psi_i\rangle~\r~\Big (A_1 |n-1\rangle+A_2 |n+1\rangle\Big ) \otimes|\psi_f\rangle
\ee
where
\bea
A_1&=&e^{-i(E_i+E_f)  \bar\tau}\int_{\tau=-\bar\tau}^{\bar\tau} d\tau (-i\alpha){1\over \sqrt{2\omega}}\sqrt{n}~e^{i\Delta E\, \tau}e^{-i\omega {\tau\over r_{R1}} }F_{\omega,k_\perp}(r_{R1})\cr
A_2&=&e^{-i(E_i+E_f) \bar \tau}\int_{\tau=-\bar\tau}^{\bar\tau} d\tau (-i\alpha){1\over \sqrt{2\omega}}\sqrt{n+1}~e^{i\Delta E\, \tau}e^{i\omega {\tau\over r_{R1}} }F_{\omega,k_\perp}(r_{R1})
\eea
where $\Delta E=E_f-E_i$. 
The probability of excitation is given by squaring the amplitudes. We have
\be
|A_1|^2={|\alpha|^2n \over \pi^2}\sinh(\pi\omega) [K_{i\omega}(k_\perp r_{R1})]^2  [{4 \sin^2[(\Delta E-{\omega\over r_{R1}})\bar\tau]\over (\Delta E-{\omega\over r_{R1}})^2}]
\label{tone}
\ee
For large $\bar\tau$ we have
\be
[{4 \sin^2[(\Delta E-{\omega\over r_{R1}})\bar\tau]\over (\Delta E-{\omega\over r_{R1}})^2}]~\r ~4\pi \bar\tau \delta(\Delta E-{\omega\over r_{R1}})
\ee
The corresponding factor in $A_2$ does not give a delta function since it describes a process where energy is not conserved; thus we ignore $A_2$. 

From (\ref{twenty}) we see that the sum over occupation numbers $n$ is weighted by $Ce^{-2\pi n\omega}$, where $C=(1-e^{-2\pi \omega})$ normalizes the sum over weights to unity. Thus the factor $n$ in (\ref{tone}) leads to the sum
\be
\sum_{n=0}^\infty Ce^{-2\pi n\omega}n={1\over e^{2\pi \omega}-1}
\ee
The sum over $k_\perp$ gives
\be
\int {d^{D-2}k_\perp\over (2\pi)^{D-2}}\r \int{\Omega_{D-3}k_\perp^{D-3} dk_\perp\over (2\pi)^{D-2}}
\ee
Thus we find that the probability of excitation under the acceleration in the interval $\tau=(-\bar\tau, \bar\tau)$ is
\bea
P&=&\int d\omega\int{\Omega_{D-3}k_\perp^{D-3} dk_\perp\over (2\pi)^{D-2}}
{|\alpha|^2 \over \pi^2}\sinh(\pi\omega) [K_{i\omega}(k_\perp r_{R1})]^2 4\pi \bar\tau \delta(\Delta E-{\omega\over r_{R1}}){1\over e^{2\pi \omega}-1}\nn
&=&{4|\alpha|^2\Omega_{D-3} \, \bar\tau\over  (2\pi)^{D-1}r_{R1}^{D-3}} e^{-\pi r_{R1} \Delta E}\int_0^\infty {x^{D-3} dx}
[K_{ir_{R1}\Delta E}(x)]^2  
\label{tthree}
\eea
Since the total proper time of acceleration was $2\bar\tau$ we have for the probability of excitation per unit proper time
\be
\Gamma={P\over 2\bar\tau}={2|\alpha|^2\Omega_{D-3} \, \over  (2\pi)^{D-1}r_{R1}^{D-3}} e^{-\pi r_{R1} \Delta E}\int_0^\infty {x^{D-3} dx}
[K_{ir_{R1}\Delta E}(x)]^2 
\label{tfive}
\ee 
We have
\be
  \int_0^\infty {x^{D-3} dx}
[K_{ir_{R1}\Delta E}(x)]^2  ={\sqrt{\pi}\Gamma[{D\over 2}-1]\Gamma[{D\over 2}-1-i r_{R1}\Delta E]\Gamma[{D\over 2}-1+ir_{R1} \Delta E ]\over 4 \Gamma[{D\over 2}-\h]}
\label{qqten}
\ee
and
\be
\Omega_{D-3}={2\pi^{{D\over 2}-1}\over \Gamma[{D\over 2}-1]}
\label{qqel}
\ee 
We can simplify the expression for $P$ for any given $D$ by using the relations \be
\Gamma[1+ix]\Gamma[1-ix]={\pi x\over \sinh(\pi x)}
, ~~~\Gamma[\h+ix]\Gamma[\h-ix]={\pi\over \cosh (\pi x)}
\label{fisix}
\ee
We find in particular
\bea
\Gamma_{D=4}&=&{|\alpha|^2 \Delta E\over 2\pi}{1\over e^{2\pi r_{R1}\Delta E}-1}\nn
\Gamma_{D=5}&=&{|\alpha|^2\, [{1\over 4}+(r_{R1}\Delta E)^2]\over 8\pi r_{R1}^2}{1\over e^{2\pi r_{R1}\Delta E}+1}
\eea
where we see the alternation of  Bose and Fermi  factors $(e^{2\pi r_{R1}\Delta E}\pm 1)$ as we change $D$ \cite{bosefermi}.

\subsubsection{Putting the factors together}

We can now collect together  all the factors which give the probability of exciting our 2-level system when it is turned around in an accelerating trajectory. The probability of excitation is given by $P=\Gamma\Delta\tau$, where $\Gamma$ is the probability of excitation per unit proper time along the path of the system and $\Delta\tau$ is the proper time for which the acceleration is maintained. $\Gamma$ is given by (\ref{tfive}), where $r_{R1}=1/a$ sets the acceleration (eq.(\ref{tsix})). The factor $\Delta\tau$ is given by eq.(\ref{tseven}), where we need to put in the value of $\bar U^r_R$ from eq.(\ref{teight}). Putting all this together, we get for the probability of excitation
\be
P={|\alpha|^2 \over 2^{D-2}\pi^{D-1\over 2}\Gamma[{D\over 2}-\h] }{e^{-\pi r_{R1}\Delta E}\Big |\Gamma[{D\over 2}-1+i r_{R1}\Delta E]\Big |^2\over r_{R1}^{D-4}}\sinh^{-1}{2 r_H/(D-3)\over \sqrt{r_{R1}^2+\bar r_R^2-{4\over (D-3)}r_{R1}r_H}}
\label{tnine}
\ee
Here we have used $U^r_R\gg 1$ in simplifying the expression for $\Delta\tau$; this inequality follows from (\ref{teight}) on noting that we are working for ${\bar r-r_H\over  r_H}\ll 1$. 
Recall that $r_H$ is  the radius of the black hole, $\Delta E$ is the energy gap in the system, and $\alpha$ measures the coupling of the system to the radiated scalar. These quantities are thus fixed for our problem.  $\bar r_R$ is the position (in the Rindler frame) upto which we agree to let the system have free fall. At $\bar r_R$ the trajectory is changed to one with an acceleration $a=1/r_{R1}$. We now have to vary $r_{R1}$ in (\ref{tnine}) till we find the value for which $P$, the excitation probability, is a minimum. 

To perform the minimization, we note a few other simplifications that follow from our limits. 

\b

(a) Consider the square root $\sqrt{r_{R1}^2+\bar r_R^2-{4\over (D-3)} r_{R1}r_H}$ that occurs in the argument of the $\sinh^{-1}$ function. The last two terms in the square root  cancel when 
\be
r_{R1}={(D-3)\over 4}{\bar r_R\over r_H}\,  \bar r_R=\sqrt{(D-3)(\bar r-r_H)\over 4r_H}\, \bar r_R\ll \bar r_R
\label{th}
\ee
where we have used (\ref{seven}) to write the Rindler frame position 
\be
\bar r_R=\sqrt{4r_H(\bar r-r_H)\over (D-3)}
\label{fione}
\ee
in terms of the original black hole coordinate $\bar r$, and in the last step we have used the fact that the point $\bar r$ is close to the horizon. Thus we see that in the square root $\sqrt{r_{R1}^2+\bar r_R^2-{4\over (D-3)} r_{R1}r_H}$  we can ignore $r_{R1}^2$. Using (\ref{fione}) in the argument of the $\sinh^{-1}$ function we write all quantities in terms of the original variables in the metric (\ref{three}) rather than Rindler frame variables\footnote{$r_{R1}$ may look like a Rindler quantity, but it is a parameter (equal to $1/a$) that we are varying over, so  there is no point in rescaling this variable.} 
\be
P={|\alpha|^2 \over 2^{D-2}\pi^{D-1\over 2}\Gamma[{D\over 2}-\h] }{e^{-\pi r_{R1}\Delta E}\Big |\Gamma[{D\over 2}-1+i r_{R1}\Delta E]\Big |^2\over r_{R1}^{D-4}}\sinh^{-1}{ \sqrt{r_H/(D-3)}\over \sqrt{(\bar r-r_H)-r_{R1}}}
\label{tnineq}
\ee

\b

(b) Note that the argument of the $\sinh^{-1}$ function satisfies
\be
{ \sqrt{r_H/ (D-3)}\over \sqrt{(\bar r-r_H)-r_{R1}}}\ge {1\over \sqrt{(D-3)}}\sqrt{r_H\over \bar r-r_H}\gg 1
\ee
where we have used the fact that we are looking at infall close to the horizon ($\bar r-r_H\ll r_H$). But for large $x$ we can write
\be
\sinh^{-1} x\approx \ln[2 x]
\label{fosix}
\ee
and so we get
\be
P={|\alpha|^2 \over 2^{D-1}\pi^{D-1\over 2}\Gamma[{D\over 2}-\h] }{e^{-\pi r_{R1}\Delta E}\Big |\Gamma[{D\over 2}-1+i r_{R1}\Delta E]\Big |^2\over r_{R1}^{D-4}}\ln\Big [{4r_H/ (D-3)\over (\bar r-r_H)-r_{R1}}\Big]
\label{tnineqq}
\ee

\b

(c) Below we will encounter the minimization of a function of the form
\be
f={1\over x^p}\ln{Q\over 1-x}, ~~~~~Q\gg 1
\label{fifour}
\ee
where $p$ is a positive number of order unity. 
 It is easy to see that such a function is minimized for $x=1-\delta$, with $\delta \ll 1$. We can expand $f$ in terms of $\ln Q, \ln\ln Q, \dots$ etc.  Working to order $\ln Q$, we find that the minimum is at $\delta\approx {1\over p\ln{Q\over \delta}}$. This gives
 \be
 f\approx \ln Q
 \label{fithree}
 \ee
 upto terms of order $\ln\ln Q$.
 
\subsection{Minimizing $P$}
 
 Finally, we can turn to the task of choosing $r_{R1}$ with the goal of minimizing the excitation probability $P$. It is helpful to consider two opposite limits of the variable $\bar r-r_H$ which determines  how close to the horizon we allow the system to fall before starting the turnaround. These limits are (i) $\bar r-r_H\ll (\Delta E)^{-1}$ and (ii) $\bar r-r_H\gg (\Delta E)^{-1}$.
 
 \subsubsection{The case $\bar r-r_H\ll (\Delta E)^{-1}$}\label{second}
 
 We imagine keeping all other parameters like $\alpha, \Delta E, r_H$ fixed while we take $\bar r-r_H\r 0$. Note that from the argument of the log in (\ref{tnineqq}) we have $r_{R1}<(\bar r-r_H)$, so we also have $r_{R1}\r 0$. (In other words, if we fall close to the hole then we need a large acceleration $a=1/r_{R1}$ to escape.) Thus we set all factors $r_{R1}\Delta E\r 0$ in
 the expression (\ref{tnineqq}) for $P$. We get
 \be
 P={C'\over x^{D-4}}\ln {Q\over 1-x}
 \ee
 where
 \be
 C'={|\alpha|^2(\Gamma[{D\over 2}-1])^2\over 2^{D-1}\pi^{D-1\over 2}\Gamma[{D\over 2}-\h]}{1\over (\bar r-r_H)^{D-4}}, ~~~Q={4 r_H\over (D-3)(\bar r-r_H)}, ~~~x={r_{R1}\over (\bar r-r_H)}
 \ee
 Since ${\bar r-r_H\over r_H}\ll 1$, we have $Q\gg1$. We can thus use (\ref{fifour}),(\ref{fithree}) for the case $D>4$ (since $p=D-4>0$). This gives
 \be
 P\approx {|\alpha|^2(\Gamma[{D\over 2}-1])^2\over 2^{D-1}\pi^{D-1\over 2}\Gamma[{D\over 2}-\h]}{1\over (\bar r-r_H)^{D-4}}\ln \Big [{4 r_H\over (D-3)(\bar r-r_H)}\Big ]
 \ee
 As $\bar r-r_H\r 0$ with other parameters fixed, we see that $P$ grows large, and so will formally exceed unity. Thus we will not be able to recover the information in the system in case we have fallen to a radius where $\bar r-r_H\ll (\Delta E)^{-1}$.
 
The case $D=4$ leads to the same conclusion. In the limit $r_{R1}\Delta E\r 0$ we have from (\ref{tnineqq})
\be
P={|\alpha|^2\over 4\pi^2}\ln \Big [{4r_H\over (\bar r-r_H)-r_{R1}}\Big ]
\ee
We see that this expression is minimized for $r_{R1}=0$, which leads to
\be
P={|\alpha|^2\over 4\pi^2}\ln \Big [{4r_H\over( \bar r-r_H)}\Big ]\gg 1
\ee
 It is interesting that in the large acceleration limit ($a=1/r_{R1}\r\infty$) the excitation probability given by $P=\Gamma\Delta \tau$ in $D=4$ goes to a constant independent of the acceleration.  See however the 
comment (c) in section \ref{comment} below: the low energy modes ($\omega\ll (\Delta \tau)^{-1}$) can give an additional diverging contribution in $D=4$, which must be regularized by the finite size of the accelerating system \cite{louko2}.

\subsubsection{The case $\bar r-r_H\gg (\Delta E)^{-1}$}
 
From the argument of the log in (\ref{tnineqq}) we see that we can let $r_{R1}\sim \bar r-r_H \gg (\Delta E)^{-1}$. In this case we have
\be
P\sim e^{-2\pi r_{R1}\Delta E}
\label{qqone}
\ee
where we have ignored all other factors since an exponential dominates above powers and logs. (Recall that we are holding parameters like $\alpha$ fixed while looking at the limit $\bar r-r_H\r 0$; one can easily redo the analysis with the full expression for $P$ if desired.)  In (\ref{qqone}) one factor $e^{-\pi r_{R1}\Delta E}$ is explicitly present in (\ref{tnineqq}), and another similar factor appears from 
$\Big |\Gamma[{D\over 2}-1+i r_{R1}\Delta E]\Big |^2$ on using (\ref{fisix}). 

Thus in the case $\bar r-r_H\gg (\Delta E)^{-1}$ we can choose an acceleration $a=1/ r_{R1}$ low enough that $P\ll 1$; i.e., the system can be returned to infinity without its information being disturbed.  Since we have from the argument of the log in (\ref{tnineqq}) that $r_{R1}<\bar r-r_H$, we see that we cannot make the exponential (\ref{qqone}) small when we reach
\be
\bar r-r_H\sim {1\over \Delta E}
\label{ththree}
\ee
Putting this together with the discussion of section \ref{second} above, we see that (\ref{ththree}) gives the point beyond which we have a probability $P\sim 1$ for the accelerated system to get excited by Unruh radiation.

\subsection{Analysis and comments}\label{comment}

Let us make a few observations about what we have found.

\b

(a) First, let us compare the condition (\ref{ththree}) with the condition 
(\ref{thfour}) that we found for the process where the infalling system tries to send information back by radiating a photon. In the computation leading to (\ref{thfour}) we did not assume any particular structure for the infalling system, which was just characterized by it total mass $m$. But let us assume for the moment that this system is also a 2-level system, with $\Delta E\sim m$.  Then we see that the conditions (\ref{ththree}) and (\ref{thfour}) give $\bar r-r_H$ of the {\it same} order. This in interesting, and something that need not have been true just on dimensional grounds -- we could have had any power of the dimensionless small number ${\bar r-r_H\over r_H}$ relating the two values of $\bar r-r_H$. 

\b

(b) The quantity $\bar r-r_H$ gives the coordinate distance to the horizon. To find the corresponding physical distance (along a constant $t$ slice) we can look at the near horizon metric in the form (\ref{eight}).  We see that the physical distance to the horizon is just $\bar r_R$, so the condition (\ref{ththree}) is
\be
d_{cr}\sim \bar r_R\sim \sqrt{r_H(\bar r-r_H)}\sim \sqrt{r_H\over \Delta E}
\label{qtwo}
\ee
where we have written $d_{cr}$ for the critical distance from the horizon  beyond which we cannot retrieve the system without an order unity probability of disturbing it by acceleration radiation. 

Let us compare $d_{cr}$ to the size $L$ of the system; after all, if $d_{cr}$ turns out to be much smaller than $L$, then the system would be essentially inside the hole at the distance $d_{cr}$, and our failure to retrieve information could be attributed to the system having crossed the horizon in some sense. But we see that $d_{cr}$ will typically be much larger than $L$. 
 Suppose the size of 2-level system was set by the uncertainty principle limit  $L\sim {1\over \Delta E}$. Then
 \be
 d_{cr}\sim \sqrt{r_H L}\gg L
 \label{qfour}
 \ee
 since the hole is much bigger than the infalling system ($r_H\gg L$). 

More generally, the system size can be even smaller than ${1\over \Delta E}$.  Consider two energy levels of an atom. The energy gap can give $(\Delta E)^{-1}\sim 6000 \, A^o$, but the size of the atom is just $\sim 1 \, A^o$. The difference between the two scales arises because the information carrying system is made of heavy objects (the electron in the atom) with strong couplings (the strength of the electric attraction), and this system size can be therefore much smaller than ${1\over \Delta E}$. Thus  (\ref{qfour}) says that a large variety of systems will suffer `effective information loss' to Unruh radiation before they touch the horizon.

\b

(c) The case $D=4$ is a little special because of the well known infra-red divergences in the computation of radiation from accelerated particles. In the computation of  bremsstrahlung, one finds that if we take a sudden change in the path of a charged particle, then there is a divergent probability of exciting the radiation field by the low energy modes. In that computation we redefine our goal: we compute the probability of exciting no radiation quanta and add it to the probability of exciting low energy quanta; this sum is well behaved in perturbation theory when we lift the infra-red cutoff. In the present problem  we are interested in the probability of entangling with the radiation field, and so we want just the probability of exciting the field.  This time the cutoff arises from the finite size of the accelerating system, something we have not taken into account. In the next section we will work with a 4+1 dimensional system where this infra-red issue does not arise.

\section{BPS brane bound states as information carrying systems}

The probability of exciting a 2-level system accelerating along a given path depends on the energy gap $\Delta E$ between the levels and the coupling between the system and the radiation field (given through $\alpha$ (eq.(\ref{thfive})). If we wish to look for estimates that might have some universal behavior, then we should  make our systems out of the degrees of freedom present in a complete theory of gravity, which we will take to be string theory. In such a theory the couplings of the system  to the radiation fields might be related to the gravitational coupling itself, and the level spacings might be related to the density of states in gravity. Thus we can hope for some universal behavior to emerge.

Let us consider a black hole in type IIB string theory, which we take with the compactification
\be
M_{9,1}\r M_{4,1}\times S^1\times T^4
\ee
The black hole we are falling into will be a Schwarzschild hole in 4+1 noncompact dimensions, with horizon radius $r_H$ (i.e., eq.(\ref{three}) with $D=5$). The $S^1$ is parametrized by a coordinate $y$ and has length  $2\pi R_y$. The $T^4$ has coordinates $w_i, i=1,\dots 4$, and volume $(2\pi)^4 V_4$. 

Now we consider how to make a system that can carry information. Consider a string wound around the $S^1$ (the fundamental string is also termed the NS1 brane).  This string can cary vibration modes along its length; these vibrations carry momentum and energy along the $S^1$. By selecting a particular vibrational excitation of the string, we choose a particular state for our system. 

Let us now see what will happen if such a string falls close to the horizon of our black hole. We will have to accelerate the string if we wish to extract it back to infinity, and it will then interact with the fields present in our theory.  The components of the graviton along the compact $T^4$ directions give scalars in $M_{4,1}$; let us focus on
\be
g_{w_1w_2}\equiv\phi
\ee
The scalar $\phi$ couples to the vibrations on the string. The transverse vibrations of the string in the directions $w_1, w_2$ are given by functions $w_1(t, y), w_2(t, y)$ along the string worldsheet. The interaction Lagrangian is
\be
S_{int}=\tilde C\int dt dy ~\phi \, [\p_+ w_1\p_- w_2 + \p_+w_2\p_-w_1]
\label{thseven}
\ee
where $\p_{\pm}=\p_{t\pm y}$. Thus the string can change its excitation state by picking up a pair of excitations -- one left moving and one right moving -- while absorbing a quantum of the $\phi$ field from the Rindler bath near the black hole horizon. This process would destroy the information that we had encoded into the string state. 

With this string theory model we do know the energy levels and couplings involved in the interaction, and can put these into our computation of the last section. But first let us analyze the nature of information storage a little further. Suppose we store infromaton on the string by creating  a left and a right moving vibration. Then, even in the absence of any acceleration, these vibrations can collide and exit their energy as a $\phi$ quantum through the interaction (\ref{thseven}). Of course {\it any} 2-level system with an energy gap $\Delta E$ suffers form the possibility that its state could spontaneously change, but if we wish to have  a reliable information storage device we should make this transition rate as small as possible. With our present system  we can achieve this by letting all the vibrations be in only one direction on the string; say purely left moving. This gives a BPS state of the string, which does not decay to any other state since it has the lowest allowed energy for the given winding number of the string and the given momentum on the string. 

Thus let the string have winding number $n_1$ around the $S^1$, and let the 
momentum carried by the vibrations be $P_y={n_p\over R_y}$. Then the different ways of partitioning this momentum among different harmonics gives $Exp[S_{2-charge}]$ states where
\be
S_{2-charge}=2\pi\sqrt{2}\sqrt{n_1n_p}
\label{theight}
\ee
But at this point we realize that this system of string winding and momentum is the same system that describes the 2-charge extremal black hole of string theory, and (\ref{theight}) is the entropy of this hole. We can get an even larger number of states by also taking $n_5$ NS5 branes wrapped on $S^1\times T^4$. The bound state of the NS5 and the strings (NS1 branes)  gives an `effective string' wrapped on the $S^1$, and adding the left moving momentum gives $Exp[S_{3-charge}]$ BPS states, with
\be
S_{3-charge}=2\pi\sqrt{n_1n_5n_p}
\label{qsix}
\ee
Thus we can take such a set of NS1-NS5-P charges and store information by choosing one of the allowed degenerate ground states of the system. Now consider the effect of accelerating such a system. The interaction of this system with the scalar $\phi$ is again known, and is in fact given by the same expression as (\ref{thseven}), since $\phi$ couples to the NS1NS5 effective string in the same way that it coupled to the elementary string. The value of the coupling $\tilde C$ was computed in \cite{dasmathur}.

Thus using such a NS1-NS5-P system as our information storage device, we find that the  acceleration can  excite the BPS ground state of the system to one of the excited states where a pair of vibrations is created, one left moving and one right moving. There are many such excitations possible, and we must sum the probabilities of creating each of them. 

At this point we find an interesting simplification: the sum of excitation probabilities we need is just the sum one computes to find the absorption cross section $\sigma$ of $\phi$ into the NS1-NS5-P system.  This cross section, in turn, has a simple form: it is just the horizon area $A_{ex}$ of the extremal black hole that has the mass and charges of the NS1-NS5-P brane system.

Let us summarize the above discussion. We wanted to make our information carrying system out of the objects present in a full theory of quantum gravity, in the hope of uncovering any universal features in the 
information loss estimates we have done. As we will discuss below, there are many different ways to store information, with different goals; one may want a large information capacity for a given energy, or one may be willing to sacrifice energy in return for stability against disturbance. Focusing on the former, it appears that a large amount of information can be stored by using the highly degenerate BPS states of string theory. The interaction rates of these systems with the scalars in the theory are known and particularly simple; after summing over levels one gets absorption/emission rates given in terms of the surface area of the extremal hole with charges $n_1, n_5, n_p$. We will use this fact to find the critical value of the approach distance $\bar r-r_H$ after which our system will not be able to return to infinity without excitation; the result will be writable in terms of the radius $r_{ex}$ of the extremal hole.

\subsection{The absorption cross section for the NS1-NS5-P system}

In this section we write the absorption cross section of the scalar $\phi$ into the 3-charge extremal system, in terms of the couplings $\alpha$ between the levels of the system and the radiation field, and the density of excited levels $\rho(E)$ that we can reach. Since the absorption cross section is known (in terms of the horizon area $A_{ex}$ of an extremal hole), we find that a certain function of $\alpha$ and $\rho(E)$ can be expressed in terms of $A_{ex}$. In the next subsection we will see that this same function of $\alpha, \rho(E)$ appears in the excitation probability when the system is accelerated.

Let us start with a  state $|\psi_i\rangle$ for the 3-charge system. We put the system in a box of volume $V$, where it interacts with the scalar field $\phi$ through the interaction Hamiltonian $g\hat O\hat\phi$. (This is a computation in flat space, there is no black hole here.) We write
\be
\hat\phi=\sum_{\vec k}[{1\over \sqrt{V}}{1\over \sqrt{2\omega}}e^{i\vec k\cdot \vec x-i\omega t}\hat a_{\vec k} + {1\over \sqrt{V}}{1\over \sqrt{2\omega}}e^{-i\vec k\cdot \vec x+i\omega t}\hat a^\dagger_{\vec k} ]
\ee
In the box we place one quantum of the field $\phi$ in the momentum mode $\vec k$; thus the energy of this mode is $\omega=|\vec k|$. The goal is to compute the probability of absorption of this $\phi$ quantum into the 3-charge system per unit time.

Let us compute the amplitude $A_{i\r f}$ for starting with the state $|\psi_i\rangle$ ar $t=-T$ and ending with the excited state   $|\psi_f\rangle$ at time $t=T$. Writing 
\be
g\langle\psi_f | \hat O |\psi_i\rangle\equiv \alpha_{i\r f}
\ee
we find
\be
A_{i\r f}={(-i\alpha_{i\r f})\over \sqrt{V}\sqrt{2\omega}}\int_{-T}^T dt e^{-i E_i (t+T)}e^{-i E_f(T-t)}e^{-i\omega t}={(-i\alpha_{i\r f})\over \sqrt{V}\sqrt{2\omega}}e^{-i(E_i+E_f)T}{2\sin[(E_f-E_i-\omega) T]\over (E_f-E_i-\omega)}
\ee
Thus
\be
|A_{i\r f}|^2={|\alpha_{i\r f}|^2\over 2\omega V }{4\sin^2[(E_f-E_i-\omega)T]\over (E_f-E_i-\omega)^2}~\r~{|\alpha_{i\r f}|^2\over 2\omega V }4\pi T\delta(E_f-E_i-\omega)
\ee
where in the second step we have taken the limit of large $T$. 

Let us now sum over the final states $|\psi_f\rangle$. We write
\be
\sum_f |\alpha_{i\r f}|^2 ~\r ~ \int dE_f \, \rho(E_f)\, |\alpha_{i\r f}|^2 ~\equiv~\int dE_f \, \beta_i(E_f)
\label{ttwo}
\ee
where $\rho(E_f)$ is the density of states around $E_f$ and we have defined 
\be
\beta_i(E_f)~\equiv~ \rho(E_f)\, |\alpha_{i\r f}|^2
\label{tfour}
\ee
The above expressions have been written as if the value of $\alpha_{i\r f}$ was the same for all final states with energy around $E_f$, but in fact the density $\rho(E_f)$ should be regarded as a `weighted density' which makes (\ref{ttwo}) true even when the $\alpha_{i\r f}$ are different for different final states around $E_f$. Putting all this together we find for the probability of absorption
\be
P=\sum_f |A_{i\r f}|^2\r \int dE_f \rho(E_f){|\alpha_{i\r f}|^2\over 2\omega V } 4\pi T         \delta(E_f-E_i-\omega)={2\pi T\over \omega V }\beta_i(E_i+\omega) 
\label{tfiveqq}
\ee

Noting that the total time of interaction was $2T$, we get for the rate of absorption
\be
\Gamma={P\over 2T}={\pi\over \omega V}\beta_i(E_i+\omega) 
\ee
Since we have just one  quantum in the box, the density of quanta is ${1\over V}$, and since the quantum is massless (and thus moves with the speed of light) the flux is ${\cal F}={1\over V}$. The absorption cross section for the massless scalar into the NS1-NS5-P system is then
\be
\sigma={\Gamma\over {\cal F}}={\pi\over \omega} \beta_i(E_i+\omega)={\pi \beta_i(E_f)\over (E_f-E_i)}
\label{tfiveq}
\ee
where we have $E_f=E_i+\omega$ from the $\delta$-function in (\ref{tfiveqq}). 
Finally, we note that the value of this cross section has been computed
(by putting in the correct couplings and density of levels) in \cite{dasmathur}. One finds that 
\be
\sigma=A_{ex}
\label{sione}
\ee
where $A_{ex}$ is the horizon area of an extremal black hole in string theory with NS1-NS5-P charges equal to $n_1, n_5, n_p$. This fact will allow us to express the information degradation of our system in terms of the geometric quantity $A_{ex}$.

\subsection{The critical distance $\bar r-r_H$ for the NS1-NS5-P system}

Let us now allow our NS1-NS5-P system to fall in the metric (\ref{three}), where now we have taken the noncompact spacetime to have dimension $D=5$. (We will however continue to write the symbol $D$ for the dimension, since the result we obtain is completely general: any brane system in any dimension whose properties agrees with the physics of black holes  would give the result (\ref{qseven}) that we find below.) 

As before we start the infall from rest, allow  the system fall to a position $\bar r$, suffer constant acceleration $a$ until it returns to $\bar r$ with its velocity reversed, and then coast back to infinity in a time reverse of the free fall part of the trajectory. 

Let the system start in the state $|\psi_i\rangle$ and consider the probability for it to get excited to a state  $|\psi_f\rangle$. Taking $\Gamma$ from (\ref{tfive}), substituting $r_R=r_{R1}$ as before, and taking the time of flight $2\bar \tau$, we have for this probability 
\be
P_{i\r f}={4|\alpha_{i\r f}|^2\Omega_{D-3} \, \bar\tau\over  (2\pi)^{D-1}r_{R1}^{D-3}} e^{-\pi r_{R1} (E_f-E_i)}\int_0^\infty {x^{D-3} dx}
[K_{ir_{R1}(E_f-E_i)}(x)]^2 
\ee 
where we have added the subscript  ${i\r f}$ to note explicitly the initial and final states. Since the initial state can transition to any final state, we must sum over final states to get the total probability
\bea
P&=&\int dE_f \rho(E_f) P_{i\r f}\nn
&=&\int d E_f \rho(E_f) {4|\alpha_{i\r f}|^2\Omega_{D-3} \, \bar\tau\over  (2\pi)^{D-1}r_{R1}^{D-3}} e^{-\pi r_{R1} (E_f-E_i)}\int_0^\infty {x^{D-3} dx}
[K_{ir_{R1}(E_f-E_i)}(x)]^2  \nn
&=&\int d E_f {4\beta_i(E_f)\Omega_{D-3} \, \bar\tau\over  (2\pi)^{D-1}r_{R1}^{D-3}} e^{-\pi r_{R1} (E_f-E_i)}\int_0^\infty {x^{D-3} dx}
[K_{ir_{R1}(E_f-E_i)}(x)]^2  \nn
&=&\int d E_f {(E_f-E_i)\sigma\over \pi}{4\Omega_{D-3} \, \bar\tau\over  (2\pi)^{D-1}r_{R1}^{D-3}} e^{-\pi r_{R1} (E_f-E_i)}\int_0^\infty {x^{D-3} dx}
[K_{ir_{R1}(E_f-E_i)}(x)]^2  \nn
\eea
where in the third step we have used (\ref{tfour}) and in the last step we have used (\ref{tfiveq}). 

We now set $\sigma=A_{ex}$ (eq.(\ref{sione})). We also choose to measure the energy in our the system starting from the BPS ground state; thus  we set $E_i=0$.  We then get
\bea
P&=&{4\Omega_{D-3}A_{ex}\bar\tau\over \pi(2\pi)^{D-1}}{1\over r_{R1}^{D-3}}\int_0^\infty d E_f E_fe^{-\pi r_{R1} E_f}\int_0^\infty {x^{D-3} dx}
[K_{ir_{R1}E_f}(x)]^2  \nn
&=&{4\Omega_{D-3}A_{ex}\bar\tau\over \pi(2\pi)^{D-1}}{1\over r_{R1}^{D-1}}\int_0^\infty d y \, y\, e^{-\pi y}\int_0^\infty {x^{D-3} dx}
[K_{iy}(x)]^2 
\eea
Let us examine the nature of this result. Since the proper time of acceleration is $2\bar\tau$ we find the probability of excitation per unit time is
\be
\Gamma={P\over 2\bar\tau} ={C_DA_{ex}\over r_{R1}^{D-1}}
\label{qqtw}
\ee
where 
\be
C_D={4\over (4\pi)^{D+1\over 2}\Gamma[{D\over 2 }-\h]}\int_0^\infty dy \, y\, e^{-\pi y} |\Gamma[{D\over 2}-1+iy]|^2
\ee
is a constant that depends only on the dimension $D$. (We have used  (\ref{qqten}),(\ref{qqel}) to simplify $C_D$.)

The probability of excitation is given by $\Gamma\Delta\tau$, with $\gamma$ given by (\ref{qqtw}) and $\Delta\tau$ given through (\ref{tseven}). Using the  approximations (\ref{teight}),(\ref{th}), we find
\be
P={2C_DA_{ex}\over r_{R1}^{D-2}}\sinh^{-1}\sqrt{r_H\over (D-3)[(\bar r - r_H)-r_{R1}]}
\ee
Using ${r_H\over \bar r-r_H}\ll 1$ we can use the approximation  (\ref{fosix}) to write
\be
P={C_DA_{ex}\over r_{R1}^{D-2}}\ln\Big [{4r_H\over (D-3)[(\bar r - r_H)-r_{R1}]}\Big ]
\ee
We have to minimize $P$ by varying $r_{R1}$. $P$ is a function of the form
\be
P={C'\over x^{D-2}}\ln {Q\over 1-x}, ~~~Q\gg 1
\ee
where we have set
\be
 C'={C_DA_{ex}\over (\bar r-r_H)^{D-2}}, ~~~x={r_{R1}\over \bar r-r_H}, ~~~ Q={4 r_H\over (D-3)(\bar r-r_H)}
 \ee
Since $Q\gg 1$, we can use (\ref{fifour}),(\ref{fithree}) to get
\be
P\approx C'\ln Q={C_DA_{ex}\over (\bar r-r_H)^{D-2}}\ln {4 r_H\over (D-3)(\bar r-r_H)}
\ee
where we have dropped double log terms $\ln\ln Q$. 
Keeping only powers of $(\bar r-r_H)$ and dropping the slower log dependence, we have
\be
P\sim \Big ({r_{ex}\over (\bar r-r_H)}\Big )^{D-2}
\ee
where $r_{ex}\sim A_{ex}^{1/(D-2)}$ is the radius of the extremal hole with charges $n_1, n_5, n_p$. The critical radius where we will effectively lose information is given by setting $P\sim 1$, which gives
\be
\bar r-r_H\sim r_{ex}
\label{qone}
\ee
The quantity $\bar r-r_H$ is the coordinate distance to the horizon. To get the physical distance (along a constant $t$ slice) we consider again the value of $\bar r_R$ as in (\ref{qtwo}), getting 
\be
d_{cr}\sim \bar r_R \sim  \sqrt{r_H(\bar r-r_H)}\sim \sqrt{r_H \, r_{ex}}
\label{qseven}
\ee
Thus we see that the critical distance from the horizon where we get effective information loss to acceleration radiation is given by the geometric mean of the radius $r_H$ of the large black hole and the radius $r_{ex}$ of the  extremal hole that would have the quantum numbers of the infalling system. As an example,  if the black hole has $r_H\sim 10^6 \, cm$ (i.e., mass of order solar mass) and the infalling system has $r_{ex}\sim 1\,  cm$, then $d_{cr}\sim 10^3 \, cm$.

\subsection{Comments}

(a) One may think that since the NS1-NS5-P system describes black holes in string theory (it is S-dual to the D1-D5-P hole studied in \cite{sv,dasmathur}), we cannot retrieve any information we put in it, even in the absence of difficulties caused by acceleration. Such is not the case, however. First, one may consider the NS1-NS5-P system at weak coupling, where it is just a  collection of strings and branes. Then we can regard  the fact that its entropy (\ref{qsix}) and absorption cross section (\ref{sione}) agree with that of a black hole as just an observation that helps us estimate the excitation probabilities in a convenient way; the brane bound state is a normal physical system like any other detector. Second, we have learnt that even if we increase the coupling to the point where the $n_1, n_5, n_p$ charges should give a black hole, we actually get a `fuzzball' rather than a traditional black hole with horizon \cite{fuzz}. A fuzzball is again just like a normal physical system, and while the information is densely packaged into its detailed structure, it is certainly extractable in principle, given that we have an arbitrary amount of time at our disposal to extract this information when we bring the system back to  infinity. 

\b

(b) In (\ref{qfour}) we had observed that $d_{cr}$ was typically much larger than the system size $L$, so the distance where we get effective information loss due to acceleration radiation cannot be confused with the distance where the system begins to physically overlap with the horizon. We can make a similar observation here, since with the system size much smaller than the black hole radius ($r_{ex}\ll r_H$) we get from (\ref{qseven})
\be
d_{cr}\gg r_{ex}
\ee
Thus the critical radius where we effectively get excited by acceleration radiation is much larger than the distance at which the extremal hole representing the system would merge with the large black hole.

\b

(c) Let us comment briefly on the NS1-NS5-P BPS bound states as an information storage device. Conventional models of information storage bits have two degenerate energy levels, with a potential barrier  in between. But for any given height of this barrier, there is still some amplitude per unit time for the state on one side to tunnel to the other side, since the actual energy eigenstates of the system are not the states on the two sides of the barrier but their symmetric and antisymmetric combinations. So if the height of the barrier is limited in some way (for example by the total energy available to make the system) then we cannot get a storage device that is accurate for all time.\footnote{See \cite{quantcomp} for some discussions on information issues in the context of quantum computation.}  If we make the two levels differ in energy by some amount, then tunneling is suppressed but the state in the higher level can spontaneously decay to the lower level by its coupling to the radiation field. On the other hand BPS states of string theory are stable in the sense that the set of such states for a system is exactly degenerate; thus there is no tunneling between states. By letting the system be in any one of $N$ degenerate states, we store an information $\ln N$ in the system.

When we are forced to accelerate the system, the BPS ground state gets excited, and we had computed $d_{cr}$ by requiring that such excitation becomes likely. By the principle of detailed balance, the excited state can transition back to one of the BPS ground states in the same order of proper time as it took for the excitation to occur, so we have a jump from one ground state to another and have lost the information we stored. But the probability of transition is not the same from any one ground state to any other; for example when we model the states of the system by the vibrations of an effective string then likely transitions are those where only a few vibration modes have their energy altered. Thus we can increase the reliability of the information storage by grouping states into sets, with states in each set differing significantly from states in another set. Such grouping would decrease $d_{cr}$ (by making it harder to confuse states), but this improvement in stability comes at a cost: now we can store less information because we have fewer `sets of states' than the number $N$ of BPS ground states. It would be interesting to develop a more detailed relation between the stability of the state, information storage capacity, and the value of $d_{cr}$. 

\section{Discussion}

We investigated the question of when information gets effectively lost as a system falls towards the horizon. We found some interesting estimates, though many more would be needed before it becomes clear if a general notion of `effective information loss' makes sense in the black hole context. We looked for the critical distance from the horizon $d_{cr}$
 beyond which the system cannot be extracted to infinity without its state being disturbed by Unruh radiation. We found that $d_{cr}\sim \sqrt{r_H\over \Delta E}$ where $r_H$ is the radius of the hole and $\Delta E$ is the energy gap between the levels of the infalling system. We also computed the critical distance by asking if the system could send its information back by a photon, and if we allow an energy budget $m\sim\Delta E$ for this system, then this critical distance agrees with the one found by the Unruh radiation condition.

We used a 2-level system with energy gap $\Delta E$  to see the effects of acceleration. It is more conventional to make information storage bits out of systems with two {\it degenerate} levels, with a potential barrier in between; the larger  the barrier height $\Delta E$, the more stable the stored information is to temperature disturbances.  We expect however that the results with such systems would be similar to results we obtained, if $\Delta E$ is of the same order in the two cases.

It has been noted that one could shield a bit carrying information from Unruh radiation by placing it in a cavity which does not allow penetration of the scalar field modes \cite{shield}. One might therefore wonder if any limit set by acceleration radiation could be bypassed by using such a cavity to shield the system. But a cavity that can provide good shielding would need to have  a large mass; a perfect shielding would presumably need {\it infinite} mass. We would need to factor in this fact into a more detailed analysis, since in our approach we are juggling three quantities: how much information is reliably stored, the energy required to store this information, and how close to the hole we are allowed to fall.

One issue we face when trying to develop general rules in our problem is that there are many ways to store information, with different goals. One does not usually think of  an information storage device as having a total energy budget $E$, but this is what we must do when we think of the system in the context of gravity. If we store information by selecting one of a given number of energy levels, then we get more information storage by having closer spaced levels. But if $\Delta E$ between levels is small, then  a lower temperature (and thus a lower acceleration) is needed to make the system jump between levels, destroying the fidelity of the information. (When $\Delta E\ll T$ excitations are enhanced because of a large number of $\phi$ quanta in the needed $\phi$ mode, de-excitations are also enhanced because of the stimulated emission into highly populated modes.) 

The BPS bound states string theory possess a large number of degenerate energy levels. In the absence of acceleration, the levels have no transitions among each other. Thus if we are not concerned about  thermal stability,  then they would be good ways to store large amounts of information. It is plausible that these brane systems are `optimal information storage devices' in some sense. For these  systems we found the critical distance  set by acceleration radiation to be given by a geometric expression $d_{cr}\sim \sqrt{r_Hr_{ex}}$, where all details about the choice of brane bound state are  encoded in $r_{ex}$, the radius of the extremal black hole with the charges of the bound state. It was not a priori obvious that we should get such a universality in $d_{cr}$, so it would be interesting to study this problem further.

The results of our estimates  suggest a change in our perspective on information in theories of gravity. First, we should think in terms of  how much information we can store for {\it given energy}. Second,  we should measure information not as a quantity contained in a given system, but in terms of how much of that information  can be reliably {\it accessed} by another observer. If the system and the observer are separated by a gravitational potential, then there is a reduction in the fidelity of the information of the system from the viewpoint of the observer, even though the system has not crossed a horizon. This degradation of information becomes infinitely strong as a horizon is approached, so we should not think of the information as lost `suddenly' when the horizon is crossed, but as a gradual loss when the horizon is approached. In fact we do not need a horizon to have  information degradation, which ties up well with the discovery in string theory that energy eigenstates do not form horizons, but rather form horizon sized quantum `fuzzballs'.

\section*{Acknowledgements}

I would like to thank James Clemens and  Perry Rice for discussions.   This  work was supported in part by DOE grant DE-FG02-91ER-40690.

\end{document}